\documentclass[twoside]{ilcws08}
\usepackage[latin1]{inputenc}
\usepackage[dvips]{graphicx,epsfig,color}
\usepackage{wrapfig,rotating}
\usepackage{amssymb,amsmath,array}

\begin{document}
\title{
Sensor/ROIC Integration using Oxide Bonding} 
\author{Zhenyu Ye \\ for the Fermilab Pixel R\&D Group
\vspace{.3cm}\\
Fermilab \\
P.O. Box 500, Batavia, Illinois - USA
}

\maketitle

\begin{abstract}
We explore the Ziptronix Direct Bond Interconnect (DBI) technology \cite{DBI} for the integration of sensors and readout integrated circuits (ROICs) for high energy physics.
The technology utilizes an oxide bond to form a robust mechanical connection between layers which serves to assist with the formation of metallic interlayer connections.
We report on testing results of sample sensors bonded to ROICs and thinned to 100 $\mu$m.
\end{abstract}

\section{Introduction}
3D integration of a multilayer stack of ICs has emerged as a solution to the demand of cost reduction and increase in functionality and performance in semi-conductor markets \cite{ibm2008,gar2008}.
The 3D integration technology includes alignment and bonding to stack the ICs, electrical interconnection between the ICs, and thinning to minimize the volume and facilitate electrical interconnections.
We have explored two 3D techniques for sensor and ROIC integration: Cu-Sn \cite{lip2008} and Ziptronix DBI bonding. 
We describe studies of the latter in these proceedings.

The Ziptronix DBI technology utilizes direct oxide bonding where initial wafer bonds are formed by placing two surfaces into contact at room temperature.
The technology includes formation of electrical interconnects integrated with the bond.
A process flow to bond pixel sensors and ROICs and make the electrical interconnects is the following. 
Patterned metal contact structures are deposited on sensor/ROIC pixels.
Silicon oxide is then deposited to cover the contact structures and wafer surface.
Chemical-mechanical polish is used to planarize the wafer surface and expose the metal contacts.
To form the initial oxide bond, wafers are carefully cleaned
and their surfaces are activated by a chemical treatment. 
They are then placed together with the metal contacts opposed. 
A final heating improves the inter-wafer bond strength and the metal-metal contact quality.
Alignment accuracies within $\pm$1 $\mu$m have been achieved.
Unlike bump-bonding where the interconnect pitch is limited by bump-to-bump shorts, the DBI technology allows very fine pitch.
Three micron pitch has been demonstrated. 
The DBI technology is also low mass.
The DBI metal contacts have a height less than 1 $\mu$m and cover a very small fraction of the surface area. 

\section{DBI integrated devices}
To demonstrate the Ziptronix DBI technology for sensor and ROIC integration, we have contracted with Ziptronix to bond sample pixel sensors to BTeV FPIX2 \cite{chr05} ROIC wafers. 
The sensor mask was designed at Fermilab and the sensors were fabricated at the MIT Lincoln Labs. 
These are 1$\times$0.7 cm$^2$, 4-side abuttable sensors with an array of 22$\times$128 p$^+$ diodes on 5 k$\Omega\cdot$cm n-bulk silicon.
The sensors employ 50 $\mu$m deep trench etch at the edge followed by doping of the sidewalls using polysilicon to provide a high quality doped surface \cite{sun05}. 
This allows the anode and thus the charge collection field to be extended to the edge without excessive leakage current. 
The FPIX2 chip was designed for use with the BTeV n+-in-n pixel detector, but can be operated with degraded performance with the opposite sign input pulse. 
When used in this mode the dynamic range of the FPIX2 amplifier is severely limited.


The DBI technology can be used either to bond wafers together or to bond dice to a wafer. 
One would normally bond known good ROIC dices to a sensor wafer. 
However, our sensor wafer layout made this problematic, so we chose instead to bond fifty sensors to each of two FPIX2 wafers.
Being conservative at the beginning of the project, the sensors after bonding were ground to a thickness of 100 $\mu$m, instead of being ground to the polysilicon trench etch.
The sensor diodes are connected to the inputs of the corresponding FPIX2 pixels by DBI metal contacts.
The trench etch filled with polysilicon is also connected by DBI metal contacts to a pad on the FPIX2 chip.
This allows biasing the sensor through the FPIX2 chip by the electrical connection between the polysilicon and the sensor backside.
Such an connection should exist either through the surface and the sensor cut edge, or by punch through 50 $\mu$m silicon bulk \cite{url}.
Bond voids corresponding to areas without bonding have been found in sixteen integrated devices by Ziptronix with scanning acoustic microscopy.
We note that according to Ziptronix, void-free bonding yields are 80-90$\%$ when processing a large number of wafers.
Excluding also known bad FPIX2 chips, we get thirty good integrated devices.
 
\section{Testing the integrated devices} 
The integrated devices were first tested by pulse injection to check the functionality of the FPIX2 chips after the bonding and thinning.
The input capacitance of the sensor to the FPIX2 chip is found to be small by comparing the measured noise level and a simulation.
The leakage currents of the sensors were measured as a function of bias voltage.
The individual electrical interconnect per pixel unit was then examined by a laser and a $\beta$ source. 
The results indicate that most of the DBI electrical interconnects between the sensor and the FPIX2 chip are formed properly.
In preparation for beam tests, the response of a device was calibrated by using a variable X-ray source.
These tests are summarized below.
\begin{figure}[t]
\hspace{0.05\columnwidth}\includegraphics[width=0.30\columnwidth]{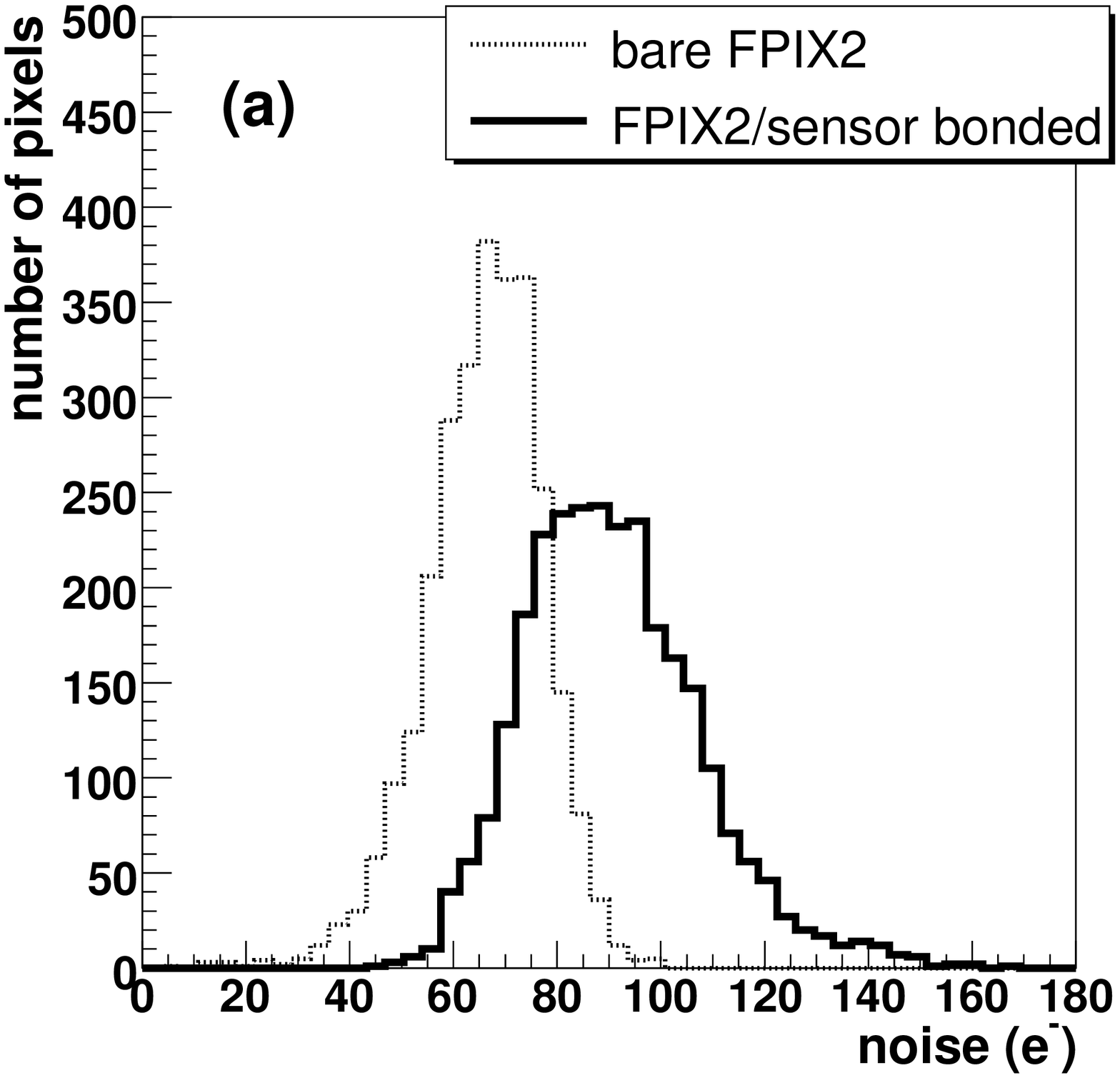} \includegraphics[width=0.30\columnwidth]{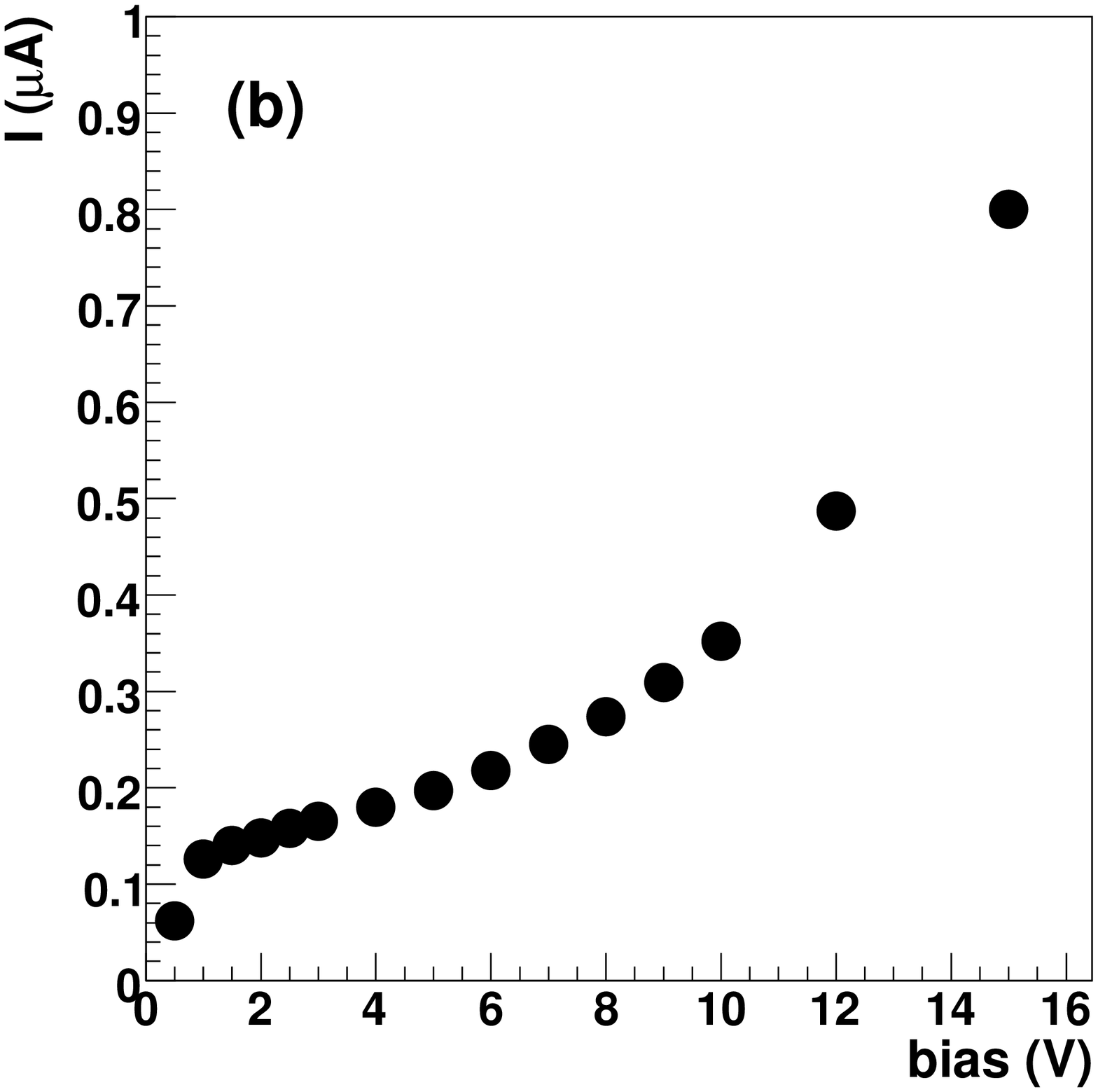} \includegraphics[width=0.30\columnwidth]{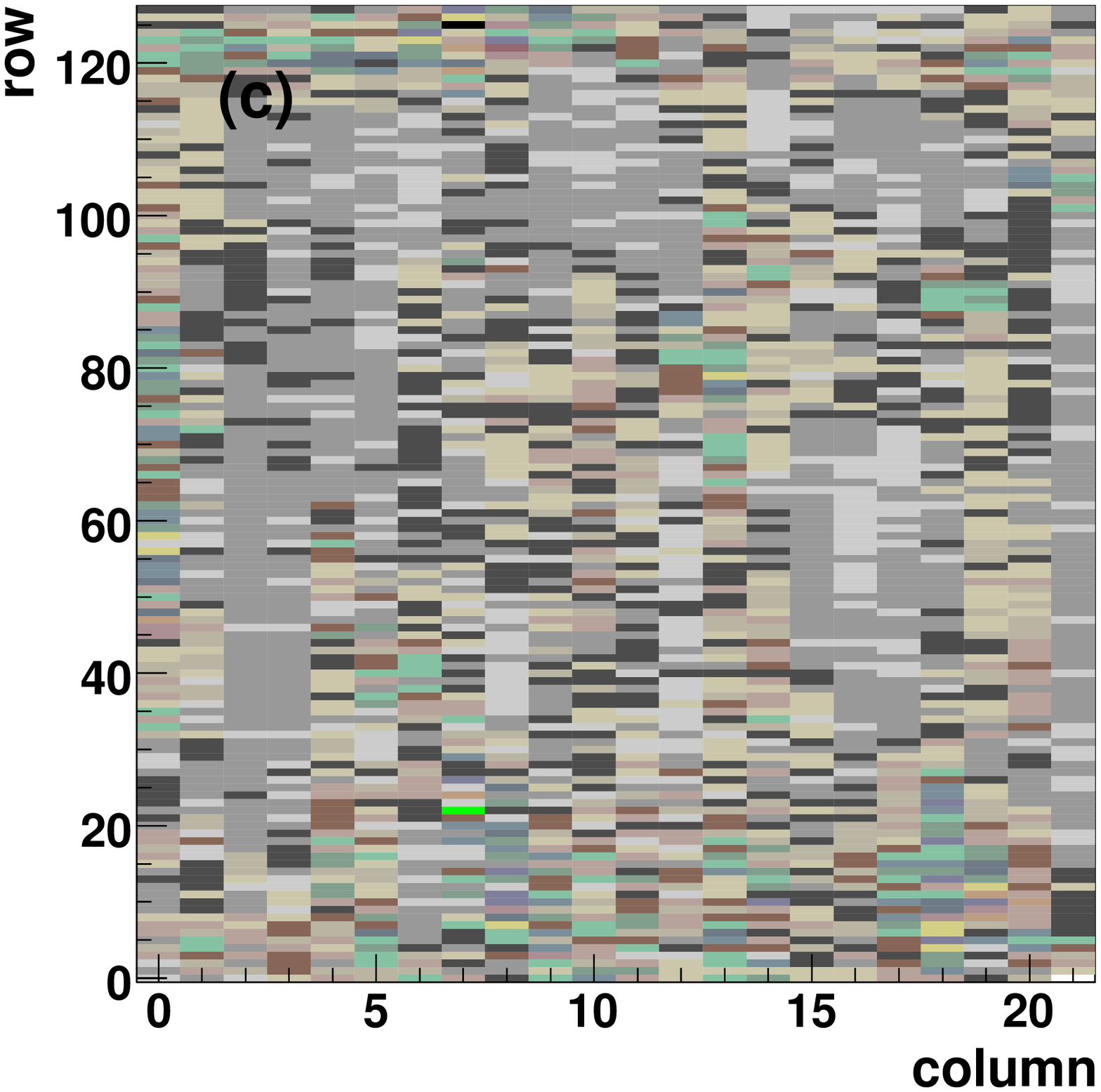}

\hspace{0.05\columnwidth}\includegraphics[width=0.30\columnwidth]{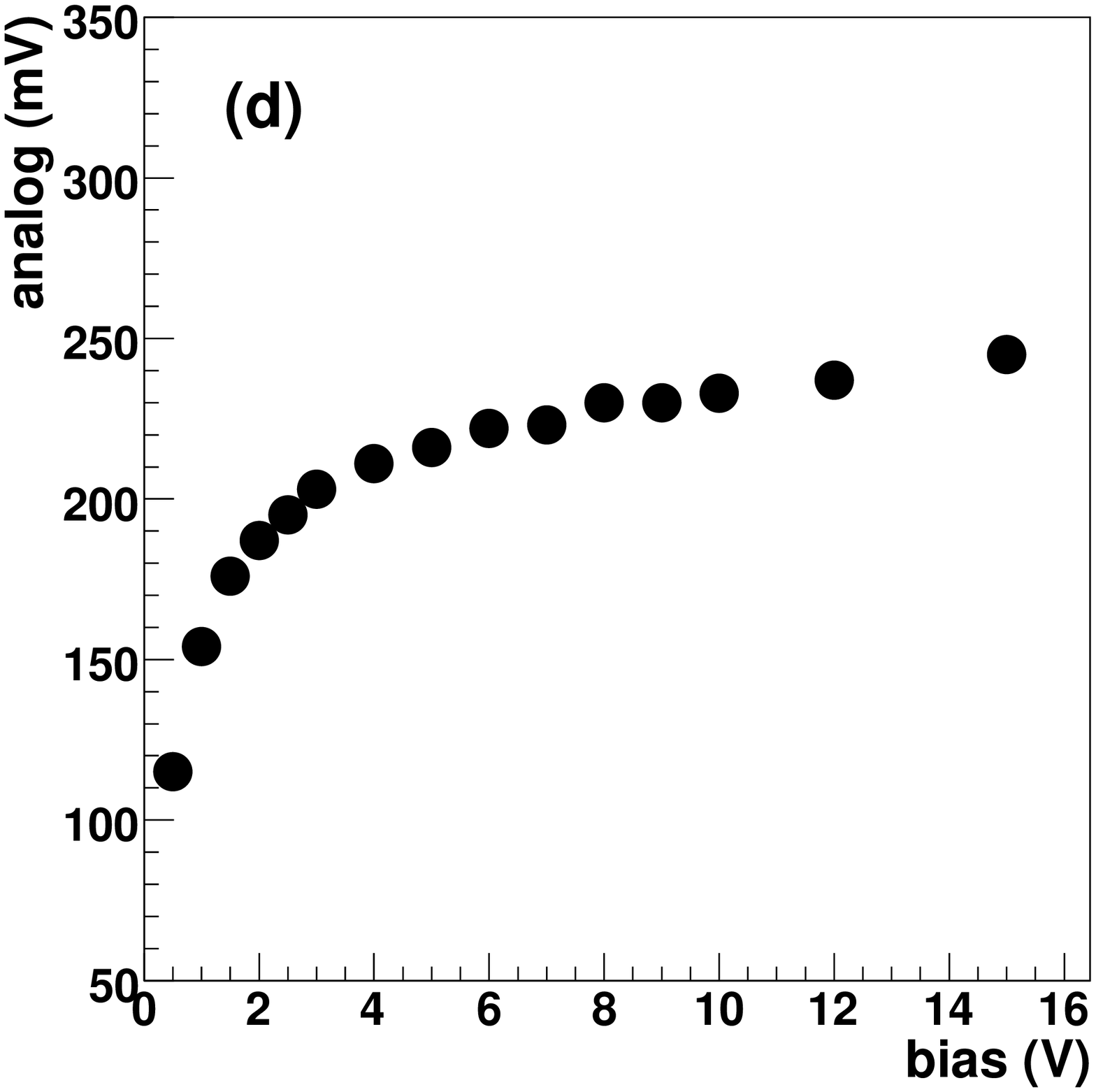} \includegraphics[width=0.30\columnwidth]{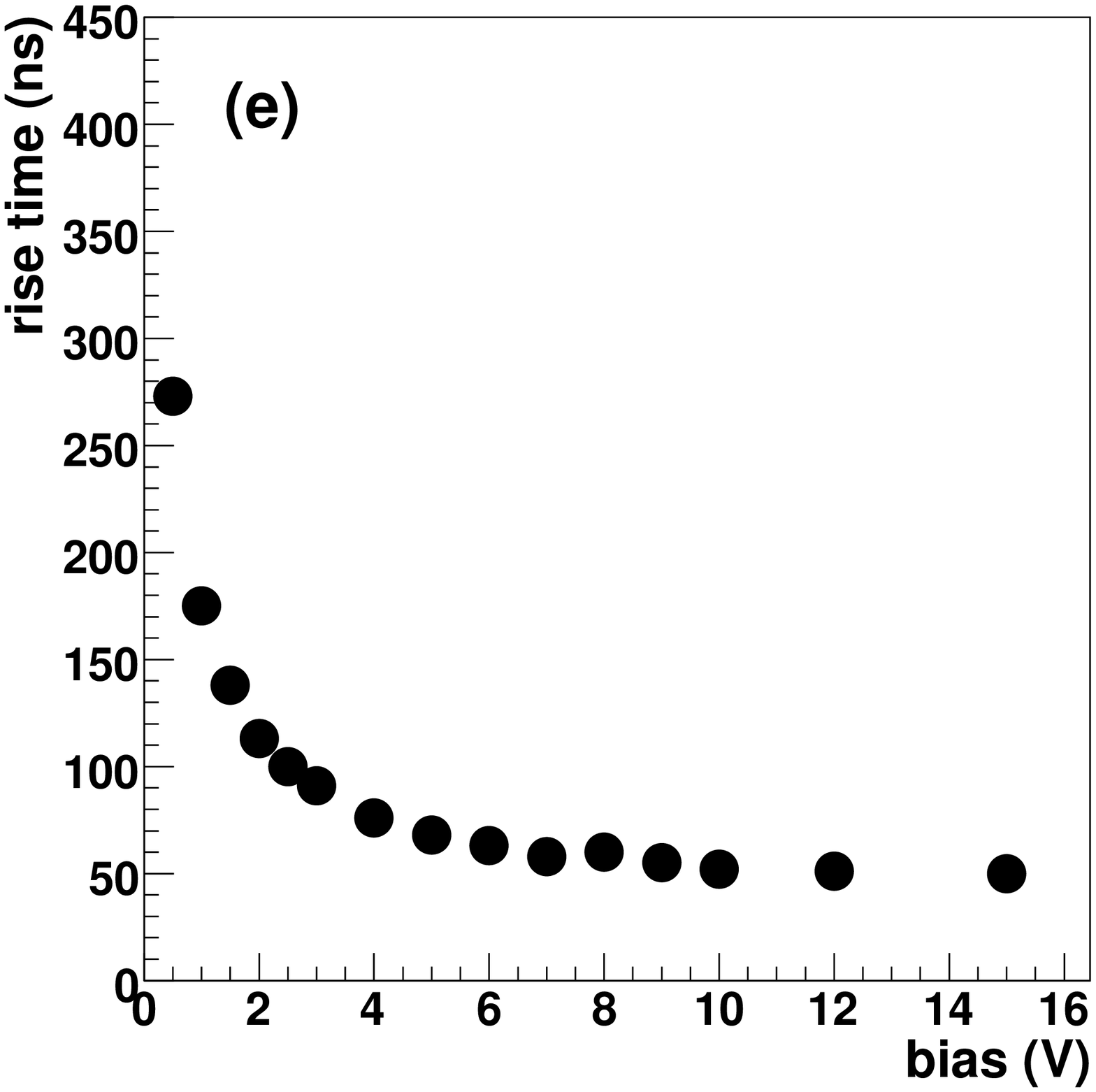} \includegraphics[width=0.30\columnwidth]{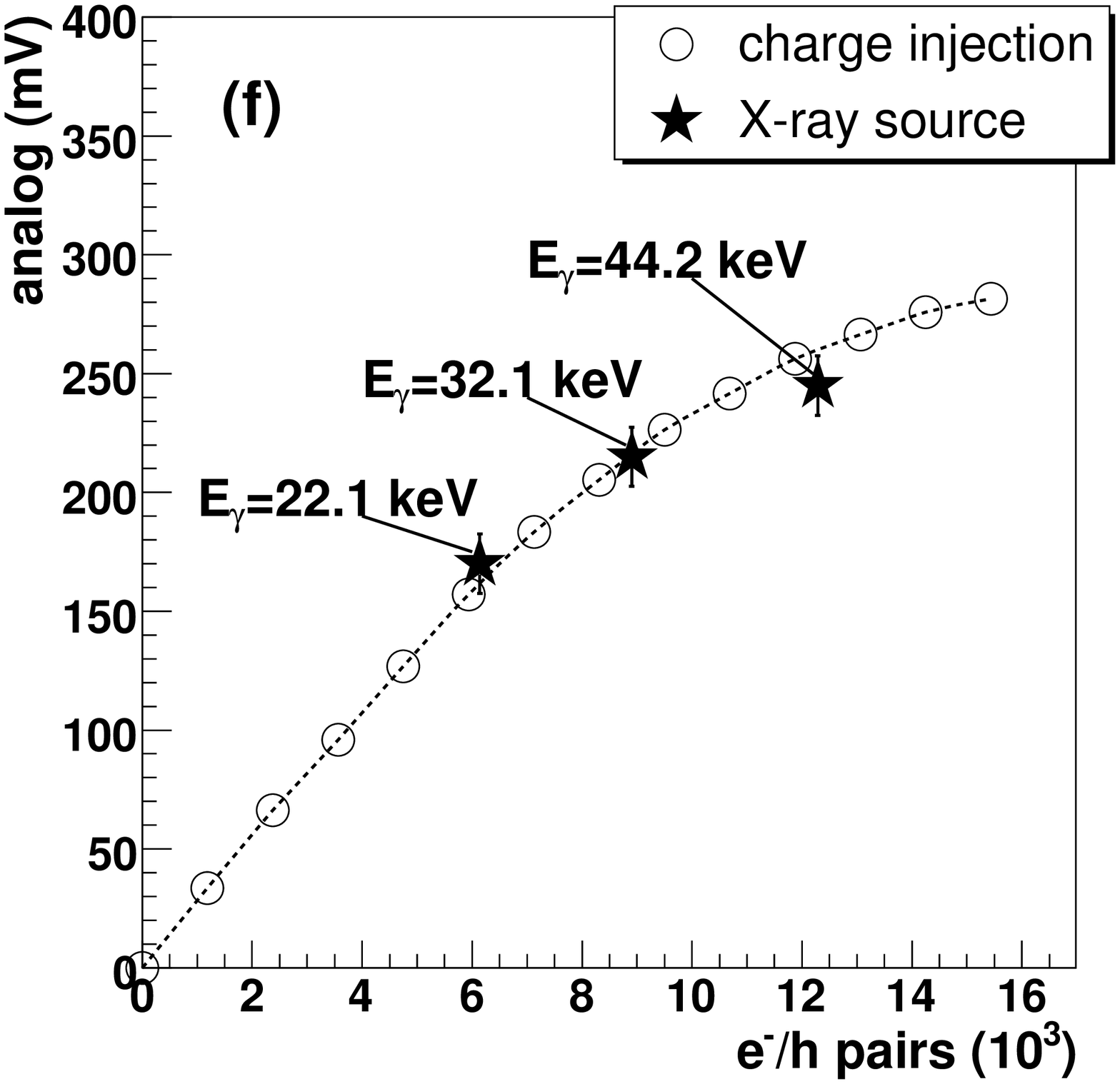}
\caption{Testing results of the DBI integrated devices:
         (a) noise distributions of FPIX2 chips measured in charge injection;
         (b) sensor leakage current as a function of the bias voltage;
         (c) hit map of an integrated device in the laser scan;
         (d) amplitude and (e) rise time of the analog amplifier output of an FPIX2 pixel unit in the laser scan as a function of the bias voltage;
         (f) amplitude of the analog amplifier output of an FPIX2 pixel unit in charge injection (open circles) and when it is exposed to a variable X-ray source (stars).
         }\label{Fig:charge}
\end{figure}

\subsection{Charge injection}
Pulses with a sharp negative step followed by a slow recovery slope were injected into the charge injection capacitors of the selected FPIX2 pixels.
The ``hit'' signals were received from the pixels whose analog outputs were above a certain threshold.
The amplitude of the pulses were scanned in a step of 1 mV.
At each step, a fixed number of pulses were injected.
Assuming Gaussian noise, the threshold and noise level of the pixels were then extracted by a cumulative Gaussian fit to the number of hits versus the amplitude of the injected pulses.
The noise distribution of a bare FPIX2 chip and one bonded to a sensor are shown in figure \ref{Fig:charge}a.
By comparing the average noise to a simulation for the FPIX2 chip, the sensor input capacitance to the FPIX2 chip is estimated to be around 0.10-0.15 pF per pixel unit.

\subsection{Sensor leakage current}
The p$^+$ diodes of the sensors are connected to the FPIX2 inputs near ground potential.
A positive bias voltage was provided through the FPIX2 chips to the trench etch filled with polysilicon.
The leakage currents of the sensors were measured as a function of the bias voltage.
A typical measurement is shown in figure \ref{Fig:charge}b.
The rapid increase of the leakage current beyond the full depletion voltage of the sensor ($\approx$8 V) is due to the bare ground backside and the associated crystal damage.
We plan to fine grind the sensors down to 50 $\mu$m, followed by backside implantation and laser annealing.
We expect that the leakage current beyond the full depletion voltage will be greatly reduced after this \cite{lip2008}.

We would like to note another way we have tested to make the sensor backside contact for the bias voltage.
In this approach, Indium-Gallium eutectic was applied and formed a very thin layer which wet the sensor backside.
The backside contact was then made through the In-Ga layer.
We tried this first on bare silicon and measured an Ohmic contact between the Indium-Gallium layer and bare silicon.
We then tried this on an integrated device with bond voids at the place where the polysilicon trench etch is connected to the FPIX2 chip.
The measured current for this device was zero when applying the bias voltage to the polysilicon trench etch.
In the approach with In-Ga, the current showed a similar behavior as figure \ref{Fig:charge}b.

\subsection{Laser scan and radiation sources}
The surfaces of the integrated sensors were scanned by a laser with single photon energy of 1.13 eV.
The hit map from one of the devices is shown in figure \ref{Fig:charge}c, in which ``hit'' signals were received from more than 99.9\% of the pixels.
The analog signal responses of the pixels near the laser spot were probed by an oscilloscope.
As can be seen in figures \ref{Fig:charge}d and \ref{Fig:charge}e, the amplitude (rise time) of these signals increases (decreases) as the bias voltage increases.
This behavior is consistent with charge collection switching from diffusion to drifting.
The integrated devices were also tested using a Sr-90 $\beta$ source.
The hit maps are consistent with efficient DBI electrical connections in all the pixel units.
All these tests indicate that the DBI interconnects between the FPIX2 chips and the sensors are properly formed.

We plan to study the charge collection efficiency profile inside each pixel unit, as well the profile near the trench in the Fermilab testbeam.
As mentioned above, the dynamic range of the FPIX2 chip is greatly reduced when it works with p$^+$-on-n detectors.
In order to know in advance whether or not the integrated devices will respond to a ``MIP'' signal, the response of an integrated device was calibrated by charge injection and a variable X-ray source.
Test pulses with a sharp positive step followed by a slow recovery slope were injected into the charge injection capacitor of a FPIX2 pixel in the device.
Debugging pads on the FPIX2 allow one access to the analog signals from one row of pixels.
The analog signal response of the pixel was measured by an oscilloscope as a function of the amplitude of the test pulses. 
The analog signal response of the same pixel was also measured when the device was exposed to a variable X-ray source.
The source is primarily an Am-241 source inceident on different metal foils, which fluoresce in the X-ray region with different energy.
The results are shown in figure \ref{Fig:charge}f, in which the amplitude of the test pulse and the X-ray energy have been converted to numbers of electron/holes.
In the conversion from the amplitude of the test pulse to charge, the nominal value of the charge injection capacitance of the FPIX2 chip is assumed.
In the conversion from the X-ray energy to charge, we assume the sensor is fully depleted with a sensitive thickness of 100 $\mu$m.
We find that for a ``MIP'' signal, the amplitude of the analog signal response is within the dynamic range of the FPIX2 chip.

\section{Conclusion}
We explore the DBI technology for the integration of pixel sensors and ROICs.
This technology provides small pitch, low mass electrical interconnects and allows thinning after bonding.
We have tested sample pixel sensors DBI-bonded with BTeV FPIX2 chips and thinned to 100 $\mu$m.
We find that the sensor input capacitance to the FPIX2 chip is small.
We also find that most if not all of the integral DBI electrical interconnections are formed properly.
We plan to study the charge collection efficiency of the sensors in beam tests.


\begin{footnotesize}

\end{footnotesize}



\begin{thebibliography}{99}
\bibitem{url} Presentation: \\ 
\verb$http://ilcagenda.linearcollider.org/contributionDisplay.py?contribId=207&sessionId=21&confId=2628$.
\bibitem{DBI} Ziptronix, Inc., 800 Perimeter Park Drive, Suite B, Morrisville, NC 27560; \\ http://www.ziptronix.com/techno/dbi.html 
\bibitem{ibm2008} IBM Journal of Research and Development Volume {\bf 52}, No. 6, 2008, Issue devoted to 3D technology.
\bibitem{gar2008} P.~Garrou, C.~Bower and P.~Ramm, {\it Handbook of 3D Integration Technology and Applications of 3D Integrated Circuits}, Wiley-VCH, 2008.
\bibitem{lip2008} R.~Lipton, {\it et~al.}, contributions to the same conference, arXiv:0901.4741.
\bibitem{chr05} D.~Christian, {\it et~al.}, Nucl. Instr. and Method {\bf A549} 165 (2005).
\bibitem{sun05} V.~Suntharalingam, {\it et~al.}, {\it Megapixel CMOS Image Sensor Fabricated in Three-dimensional Integrated Circuit Technology}, IEEE SSCC 2005, pp356-7.
\end{thebibliography}
\end{document}